# Symmetric Allocations for Distributed Storage

Derek Leong
Department of Electrical Engineering
California Institute of Technology
Pasadena, California 91125, USA
derekleong@caltech.edu

Alexandros G. Dimakis
Department of Electrical Engineering
University of Southern California
Los Angeles, California 90089, USA
dimakis@usc.edu

Tracey Ho
Department of Electrical Engineering
California Institute of Technology
Pasadena, California 91125, USA
tho@caltech.edu

*Abstract*—We consider the problem of optimally allocating a given total storage budget in a distributed storage system. A source has a data object which it can code and store over a set of storage nodes; it is allowed to store any amount of coded data in each node, as long as the total amount of storage used does not exceed the given budget. A data collector subsequently attempts to recover the original data object by accessing each of the nodes independently with some constant probability. By using an appropriate code, successful recovery occurs when the total amount of data in the accessed nodes is at least the size of the original data object. The goal is to find an optimal storage allocation that maximizes the probability of successful recovery. This optimization problem is challenging because of its discrete nature and nonconvexity, despite its simple formulation. *Symmetric* allocations (in which all nonempty nodes store the same amount of data), though intuitive, may be suboptimal; the problem is nontrivial even if we optimize over only symmetric allocations. Our main result shows that the symmetric allocation that spreads the budget maximally over all nodes is asymptotically optimal in a regime of interest. Specifically, we derive an upper bound for the suboptimality of this allocation and show that the performance gap vanishes asymptotically in the specified regime. Further, we explicitly find the optimal *symmetric* allocation for a variety of cases. Our results can be applied to distributed storage systems and other problems dealing with reliability under uncertainty, including delay tolerant networks (DTNs) and content delivery networks (CDNs).

## I. INTRODUCTION

Consider a distributed storage system comprising $n$ storage nodes. A source has a data object of unit size which is to be coded and stored in a distributed manner over these nodes; it could, for instance, split the data object into multiple chunks and then replicate them redundantly over the nodes. Let $x_i$ be the amount of data stored in node $i \in \{1, \ldots, n\}$. Any amount of data may be stored in each node, as long as the total amount of storage used is at most a given budget $T$, that is, $\sum_{i=1}^{n} x_i \leq T$. This is a realistic constraint if there is limited transmission bandwidth or storage space, or if it is too costly to mirror the data object in its entirety in every node. At some time after the creation of this coded storage, a data collector attempts to recover the original data object by accessing only the data stored in a random subset $\mathbf{r}$ of the nodes, where $\mathbf{r}$ is

This work has been supported in part by the Air Force Office of Scientific Research under grant FA9550-10-1-0166 and Caltech's Lee Center for Advanced Networking.

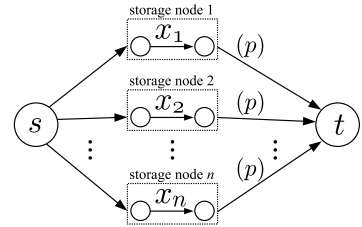

Fig. 1. Information flows in a distributed storage system. The source $s$ has a data object of unit size which it can code and store over $n$ storage nodes. Subsequently, a data collector $t$ attempts to recover the original data object by accessing each of the $n$ nodes independently with probability $p$.

to be specified by the assumed access model or failure model (nodes or links may fail probabilistically, for example).

By using a good coding scheme that enables successful recovery whenever the total amount of data accessed by the data collector is at least the size of the original data object, we can decouple the problems of (i) allocating the given budget among the nodes, that is, determining the values of $x_1, \ldots, x_n$, and (ii) designing a coding scheme for such an allocation. (This can be achieved with a suitable MDS code, or with random linear codes, for example.) Consequently, the probability of successful recovery for an allocation $\{x_1, \ldots, x_n\}$ can be written as

$$\mathbb{P}\left[\text{successful recovery}\right] = \mathbb{P}\left[\sum_{i \in \mathbf{r}} x_i \geq 1\right].$$

Our goal is to find an optimal allocation that maximizes this recovery probability, subject to the given budget constraint.

In this paper, we assume a natural access model in which the data collector accesses each of the $n$ storage nodes independently with probability $p$, as depicted in Fig. 1. In other words, each node $i$ appears in subset $\mathbf{r}$ independently with probability $p$. The resulting problem can be interpreted as that of maximizing the reliability of data storage in a system where each node fails independently with probability $(1-p)$. It turns out that this is a challenging nonconvex optimization problem, despite the simplicity of its formulation. The problem was investigated by several people at UC Berkeley [1], and has led to recent work on distributed storage allocation (see, for e.g., [2]–[5]). The reader is encouraged to work out some small examples to understand where the complexity of the problem lies. One may expect to always find an optimal allocation that is *symmetric*, i.e. with all nonzero $x_i$ being equal, but this intuition is incorrect. For instance, the following counterexample shows that symmetric allocations can be suboptimal: Given $(n, p, T) = \left(5, \frac{2}{3}, \frac{7}{3}\right)$, the nonsymmetric allocation $\left\{\frac{2}{3}, \frac{2}{3}, \frac{1}{3}, \frac{1}{3}, \frac{1}{3}\right\}$ yields a recovery probability of $0.90535$, which is strictly greater than the recovery probabilities for the five symmetric allocations, of which $\left\{\frac{7}{6}, \frac{7}{6}, 0, 0, 0\right\}$ and $\left\{\frac{7}{12}, \frac{7}{12}, \frac{7}{12}, \frac{7}{12}, 0\right\}$ achieve the highest recovery probability of $0.88889$. In this case, maximal spreading of the budget over all nodes, i.e. assigning $x_i = \frac{T}{n}$ for all $i$, turns out to perform poorly, even though one may expect greater reliability from "spreading eggs over multiple baskets."

**Our Contribution:** In this paper, we show that the intuitive symmetric allocation that spreads the budget maximally over all nodes is indeed asymptotically optimal in a regime of interest. Specifically, we derive an upper bound for the suboptimality of this allocation, and show that the performance gap vanishes asymptotically as the total number of storage nodes $n$ grows, when $T > \frac{1}{p}$. This is a regime of interest because a high probability of successful recovery is possible when $T > \frac{1}{p} \iff pT > 1$: The expected total amount of data accessed by the data collector is given by

$$\mathbb{E}\left[\sum_{i=1}^{n} x_i Y_i\right] = \sum_{i=1}^{n} x_i \mathbb{E}[Y_i] = p \sum_{i=1}^{n} x_i \leq pT,$$

where $Y_i$'s are independent Bernoulli$(p)$ random variables. Therefore, the data collector would be able to access a sufficient amount of data *in expectation* for successful recovery if $pT > 1$. In addition, we explicitly find the optimal *symmetric* allocation for a wide range of parameter values of $p$ and $T$.

**Related Work:** Jain et al. [2] evaluated the performance of symmetric allocations experimentally in the context of routing in a delay tolerant network (DTN). The authors also presented an alternative formulation using Gaussian distributions to model partial access to nodes. Note that the related theoretical claims found in [2] and its associated technical report contain some proofs that are incomplete and partially inaccurate. In [3]–[5], a different access model was considered in which the data collector accesses a random *fixed-size* subset of nodes. Various storage allocation problems have also been studied in a *nonprobabilistic* setting, with the objective of minimizing the total storage budget required to satisfy a given set of recovery requirements in a network (see, for e.g., [6], [7]).

In the next section, we define the problem formally and state our main results, which are then proved in the following section.

## II. PROBLEM DEFINITION AND MAIN RESULTS

We adopt the following notation throughout the paper:

- $n$    total number of storage nodes, $n \geq 2$
- $p$    access probability, $0 < p < 1$
- $x_i$    amount of data stored in storage node $i$, $x_i \geq 0$, where $i \in \{1, \ldots, n\}$
- $T$    total storage budget, $1 \leq T \leq n$

Allocations are expressed as multisets, e.g. $\{1, 1, 0, 0\}$, and we write "$\mathcal{B}(n, p)$" as shorthand for the binomial random variable with $n$ trials and success probability $p$.

We consider the storage allocation problem where the data collector accesses each of the $n$ storage nodes independently with probability $p$; successful recovery occurs iff the total amount of data stored in the accessed nodes is at least $1$. We seek an optimal allocation $\{x_1, \ldots, x_n\}$, among all allocations of the budget $T$, that maximizes the probability of successful recovery for a given choice of $n$, $p$, and $T$. This optimization problem can be expressed as follows:

$\mathbf{\Pi}(n, p, T):$

$$\underset{x_1, \ldots, x_n}{\text{maximize}} \sum_{\mathbf{r} \in \mathcal{P}(\{1, \ldots, n\})} p^{|\mathbf{r}|}(1-p)^{n-|\mathbf{r}|} \cdot \mathbf{I}\left[\sum_{i \in \mathbf{r}} x_i \geq 1\right]$$

subject to
$$\sum_{i=1}^{n} x_i \leq T$$
$$x_i \geq 0 \quad \forall\ i \in \{1, \ldots, n\},$$

where $\mathcal{P}(S)$ denotes the power set of $S$, and $\mathbf{I}[G] = 1$ if statement $G$ is true, and $0$ otherwise. For the trivial budget $T = 1$, the optimal allocation is $\{1, 0, \ldots, 0\}$; for $T = n$, the optimal allocation is $\{1, \ldots, 1\}$. The problem is difficult in general because the objective function is discrete and nonconvex, and there is a large space of feasible allocations to consider.

Let $\bar{\mathbf{x}}(n, T, m)$ be the *symmetric* allocation for $n$ nodes that uses a total storage of $T$ and contains exactly $m \in \{1, 2, \ldots, n\}$ nonempty nodes, that is,

$$\bar{\mathbf{x}}(n, T, m) \triangleq \left\{\underbrace{\frac{T}{m}, \ldots, \frac{T}{m}}_{m \text{ terms}}, \underbrace{0, \ldots, 0}_{(n-m) \text{ terms}}\right\}.$$

Our first result bounds the suboptimality of the symmetric allocation $\bar{\mathbf{x}}(n, T, m{=}n)$, and shows that its recovery probability approaches that of an optimal allocation as $n$ goes to infinity when $T > \frac{1}{p}$:

**Theorem 1.** *The gap between the probabilities of successful recovery for an optimal allocation and for the symmetric allocation $\bar{\mathbf{x}}(n, T, m{=}n)$ is at most*

$$pT\, \mathbb{P}\left[\mathcal{B}(n-1, p) \leq \left\lceil\frac{n}{T}\right\rceil - 2\right].$$

*If $p$ and $T$ are fixed such that $T > \frac{1}{p}$, then this gap approaches zero as $n$ goes to infinity.*

The regime $T > \frac{1}{p}$ is of interest because the recovery probability would be bounded away from $1$ if $T < \frac{1}{p} \iff pT < 1$ instead. This follows from the application of Markov's inequality to the random variable $W$ denoting the total amount of data accessed by the data collector, which gives $\mathbb{P}[W \geq 1] \leq \mathbb{E}[W]$. Since $\mathbb{P}[W \geq 1]$ is just the probability of successful recovery, and $\mathbb{E}[W] \leq pT$ as shown in the introduction, we have

$$\mathbb{P}[\text{successful recovery}] \leq pT.$$

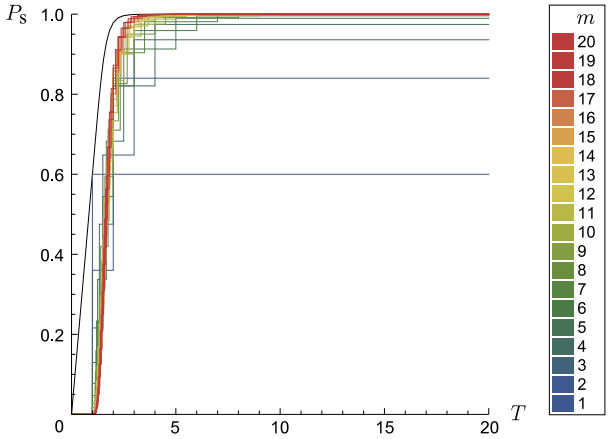

Fig. 2. Plot of probability of successful recovery $P_S$ against budget $T$ for each symmetric allocation $\bar{\mathbf{x}}(n, T, m)$, for $(n, p) = (20, \frac{3}{5})$. Parameter $m$ denotes the number of nonempty nodes in the symmetric allocation. The black curve gives an upper bound for the recovery probability of an optimal allocation, as derived in Lemma 1.

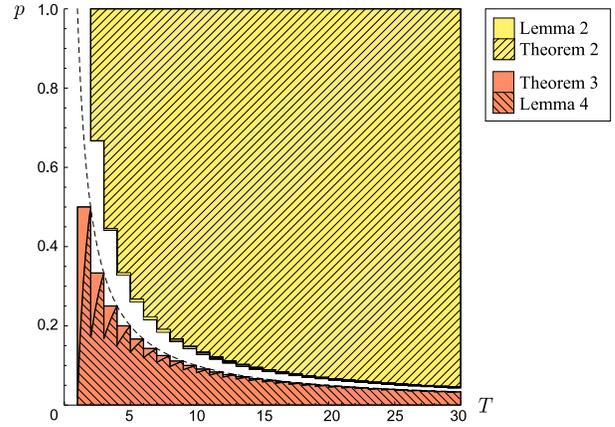

Fig. 3. Plot of access probability $p$ against budget $T$, showing regions of $(T, p)$ over which the sufficient conditions of the theorems are satisfied. The black dashed curve marks the points satisfying $p = \frac{1}{T}$. "Maximal spreading" is optimal among symmetric allocations in the colored regions above the curve, while "minimal spreading" is optimal among symmetric allocations in the colored regions below the curve.

The rest of our results deal with the optimization problem restricted over *symmetric* allocations. The problem appears nontrivial despite this simplification, as demonstrated by Fig. 2 which compares the performance of different symmetric allocations over different budgets, for a particular choice of $n$ and $p$; the value of $m$ corresponding to the optimal symmetric allocation can change drastically with varying budget.

Fortunately, as we shall see in the following section, the recovery probability for a symmetric allocation can be expressed as the tail probability of a binomial distribution. This facilitates analysis and enables us to provide a sufficient condition for "*maximal* spreading" to be optimal among symmetric allocations (Theorem 2), and for "*minimal* spreading" to be optimal among symmetric allocations (Theorem 3):

**Theorem 2.** *If* $T \geq \left\lceil \frac{4}{3p} \right\rceil$, *then* $\bar{\mathbf{x}}(n, T, m = \lfloor \lfloor \frac{n}{T} \rfloor T \rfloor)$ *or* $\bar{\mathbf{x}}(n, T, m = n)$ *is an optimal symmetric allocation. Both candidate allocations are identical when* $\frac{n}{T} \in \mathbb{Z}^+$, *i.e.* $T = n, \frac{n}{2}, \frac{n}{3}, \ldots$.

**Theorem 3.** *If* $T \leq \left\lfloor \frac{1}{p} \right\rfloor$, *then* $\bar{\mathbf{x}}(n, T, m = \lfloor T \rfloor)$ *is an optimal symmetric allocation.*

Fig. 3 summarizes these theorems in the form of a region plot. Our results cover all choices of $p$ and $T$ except for the gap around $p = \frac{1}{T}$, which diminishes with increasing $T$. "Minimal spreading" and "maximal spreading" may both be suboptimal among symmetric allocations in this gap; for example, $\bar{\mathbf{x}}(n, T, m = \lfloor 2T \rfloor)$ and $\bar{\mathbf{x}}(n, T, m = \lfloor 3T \rfloor)$ are the optimal symmetric allocations for $(n, p, T) = \left(10, \frac{9}{25}, \frac{5}{2}\right)$ and $\left(10, \frac{3}{5}, \frac{12}{5}\right)$, respectively. In general, for any access probability $p$, the optimal symmetric allocation changes from "minimal spreading" to "maximal spreading" eventually, as budget $T$ increases. This transition, which is not necessarily sharp, appears to occur at around $T = \frac{1}{p}$. Interestingly, when $T = \frac{1}{p}$ exactly, we observe numerically that $\bar{\mathbf{x}}(n, T, m = \lfloor T \rfloor)$ is the optimal symmetric allocation for *most* values of $T$; the optimal symmetric allocation changes continuously over the intervals $1.5 \leq T < 2$ and $2.5 \leq T \leq 2.8911$, while $\bar{\mathbf{x}}(n, T, m = \lfloor 2T \rfloor)$ is optimal for $3.5 \leq T \leq 3.5694$. These findings suggest that it may be difficult to specify an optimal symmetric allocation for values of $p$ and $T$ in the gap; we can, however, restrict our search for an optimal symmetric allocation to $\left\lceil \frac{n}{T} \right\rceil$ candidates, as explained in the next section.

## III. ANALYSIS

To prove Theorem 1, we need an upper bound for the probability of successful recovery for an optimal allocation (over all symmetric and nonsymmetric allocations):

**Lemma 1.** *The probability of successful recovery for an optimal allocation is at most*

$$\sum_{r=0}^{n} \min\left(\frac{rT}{n}, 1\right) \mathbb{P}[\mathcal{B}(n, p) = r].$$

*Proof of Lemma 1:* Consider a feasible allocation $\{x_1, \ldots, x_n\}$; we have $\sum_{i=1}^{n} x_i \leq T$, where $x_i \geq 0$, $i = 1, \ldots, n$. Let $S_r$ denote the number of $r$-subsets of $\{x_1, \ldots, x_n\}$ that have a sum of at least 1, where $r \in \{1, \ldots, n\}$. By conditioning on the number of nodes accessed by the data collector, the probability of successful recovery for this allocation can be written as

$\mathbb{P}[\text{successful recovery}]$

$$= \sum_{r=1}^{n} \mathbb{P}[\text{successful recovery} \mid \text{exactly } r \text{ nodes were accessed}]$$
$$\qquad \cdot \mathbb{P}[\text{exactly } r \text{ nodes were accessed}]$$
$$= \sum_{r=1}^{n} \frac{S_r}{\binom{n}{r}} \cdot \mathbb{P}[\mathcal{B}(n, p) = r]. \qquad (1)$$

We proceed to find an upper bound for $S_r$. For a given $r$, we can write $S_r$ inequalities of the form $x'_1 + \cdots + x'_r \geq 1$. Summing up these $S_r$ inequalities produces an inequality of the form $a_1 x_1 + \cdots + a_n x_n \geq S_r$. Since each $x_i$ belongs to exactly $\binom{n-1}{r-1}$ distinct $r$-subsets of $\{x_1,\ldots,x_n\}$, it follows that $0 \leq a_i \leq \binom{n-1}{r-1}$, $i=1,\ldots,n$. Therefore,

$$S_r \leq a_1 x_1 + \cdots + a_n x_n \leq \binom{n-1}{r-1} \sum_{i=1}^{n} x_i \leq \binom{n-1}{r-1} T.$$

Since $S_r$ is also at most $\binom{n}{r}$, i.e. the total number of $r$-subsets, we have $S_r \leq \min\left(\binom{n-1}{r-1}T, \binom{n}{r}\right)$. Substituting this bound into (1) completes the proof. ∎

*Proof of Theorem 1:* The probability of successful recovery for the symmetric allocation $\bar{\mathbf{x}}(n,T,m{=}n)$ is the probability of accessing at least $\lceil 1/(\frac{T}{n}) \rceil = \lceil \frac{n}{T} \rceil$ nodes, which is $\sum_{r=\lceil \frac{n}{T} \rceil}^{n} \mathbb{P}[\mathcal{B}(n,p) = r]$. The suboptimality gap for this allocation is therefore at most the difference between its recovery probability and the upper bound of Lemma 1, which is given by

$$\sum_{r=1}^{\lceil \frac{n}{T} \rceil - 1} \frac{rT}{n} \binom{n}{r} p^r (1-p)^{n-r}$$
$$= pT \sum_{\ell=0}^{\lceil \frac{n}{T} \rceil - 2} \binom{n-1}{\ell} p^\ell (1-p)^{(n-1)-\ell}$$
$$= pT \, \mathbb{P}\left[\mathcal{B}(n-1,p) \leq \lceil \tfrac{n}{T} \rceil - 2\right] \triangleq \delta(n,p,T),$$

as required. Assuming now that $T > \frac{1}{p}$, we have

$$\delta(n,p,T) \leq pT \, \mathbb{P}\left[\mathcal{B}(n-1,p) \leq \frac{n-1}{T}\right], \text{ since } \begin{array}{l}\lceil \frac{n}{T} \rceil - 2 \\ < \frac{n}{T} + 1 - 2 \\ < \frac{n}{T} - \frac{1}{T}\end{array}$$
$$= pT \, \mathbb{P}\left[\mathcal{B}(n-1,p) \leq \frac{1}{pT}(n-1)p\right]$$
$$\leq pT \, \exp\left(-\frac{(n-1)p}{2}\left(1 - \frac{1}{pT}\right)^2\right). \quad (2)$$

Inequality (2) follows from the observation that $\frac{1}{pT} \in (0,1)$, and the subsequent application of the Chernoff bound for deviation below the mean of the binomial distribution (see, for e.g., [8]). For fixed $p$ and $T$, this upper bound approaches zero as $n$ goes to infinity. ∎

Before proceeding with the proofs on the optimal symmetric allocation, we make a number of important observations about symmetric allocations in general. Successful recovery for the symmetric allocation $\bar{\mathbf{x}}(n,T,m)$ occurs iff at least $\lceil 1/(\frac{T}{m}) \rceil = \lceil \frac{m}{T} \rceil$ nonempty nodes are accessed. Therefore, the corresponding probability of successful recovery can be written as

$$P_{\mathrm{S}}(p,T,m) \triangleq \mathbb{P}\left[\mathcal{B}(m,p) \geq \left\lceil \tfrac{m}{T} \right\rceil\right].$$

Given $n$, $p$, and $T$, we have $\lceil \frac{m}{T} \rceil = k$ when $m \in ((k-1)T, kT]$, for $k = 1, 2, \ldots, \lfloor \frac{n}{T} \rfloor$, and finally, $\lceil \frac{m}{T} \rceil = \lfloor \frac{n}{T} \rfloor + 1$ when $m \in (\lfloor \frac{n}{T} \rfloor T, n]$. Since $\mathbb{P}[\mathcal{B}(m,p) \geq k]$ is nondecreasing in $m$ for constant $p$ and $k$, it follows that $P_{\mathrm{S}}(p,T,m)$ is maximized over each of these intervals of $m$ when we pick $m$ to be the largest integer in the corresponding interval. Thus, given $n$, $p$, and $T$, we can find an optimal $m^*$ that maximizes $P_{\mathrm{S}}(p,T,m)$ over all $m$ from among $\lceil \frac{n}{T} \rceil$ candidates:

$$\{\lfloor T \rfloor, \lfloor 2T \rfloor, \ldots, \lfloor \lfloor \tfrac{n}{T} \rfloor T \rfloor, n\}. \quad (3)$$

For $m = \lfloor kT \rfloor$, where $k \in \mathbb{Z}^+$, the corresponding probability of successful recovery is given by $P_{\mathrm{S}}(p,T,m{=}\lfloor kT \rfloor) = \mathbb{P}[\mathcal{B}(\lfloor kT \rfloor, p) \geq k]$. The difference between the probabilities of successful recovery for consecutive values of $k \in \mathbb{Z}^+$ can be written as

$$\Delta(p,T,k) \triangleq P_{\mathrm{S}}(p,T,m{=}\lfloor (k+1)T \rfloor) - P_{\mathrm{S}}(p,T,m{=}\lfloor kT \rfloor)$$
$$= \mathbb{P}[\mathcal{B}(\lfloor (k+1)T \rfloor, p) \geq k+1] - \mathbb{P}[\mathcal{B}(\lfloor kT \rfloor, p) \geq k]$$
$$= \sum_{i=1}^{\min(\alpha_{k,T}-1, k)} \mathbb{P}[\mathcal{B}(\lfloor kT \rfloor, p) = k - i] \cdot \mathbb{P}[\mathcal{B}(\alpha_{k,T}, p) \geq i+1]$$
$$- \mathbb{P}[\mathcal{B}(\lfloor kT \rfloor, p) = k] \cdot \mathbb{P}[\mathcal{B}(\alpha_{k,T}, p) = 0],$$

where $\alpha_{k,T} \triangleq \lfloor (k+1)T \rfloor - \lfloor kT \rfloor$. The above expression is obtained by comparing the branches of the probability tree for $\lfloor kT \rfloor$ vs $\lfloor (k+1)T \rfloor$ independent Bernoulli trials: the first term describes unsuccessful events ("$\mathcal{B}(\lfloor kT \rfloor, p) < k$") becoming successful ("$\mathcal{B}(\lfloor (k+1)T \rfloor, p) \geq k+1$") after the additional $\alpha_{k,T}$ trials, while the second term describes successful events ("$\mathcal{B}(\lfloor kT \rfloor, p) \geq k$") becoming unsuccessful ("$\mathcal{B}(\lfloor (k+1)T \rfloor, p) < k+1$") after the additional $\alpha_{k,T}$ trials. After further simplification, we arrive at

$$\Delta(p,T,k) = p^k (1-p)^{\lfloor (k+1)T \rfloor - k}$$
$$\cdot \left\{\sum_{i=1}^{\min(\alpha_{k,T}-1,k)} \sum_{j=i+1}^{\alpha_{k,T}} \binom{\lfloor kT \rfloor}{k-i}\binom{\alpha_{k,T}}{j}\left(\frac{p}{1-p}\right)^{-i+j} - \binom{\lfloor kT \rfloor}{k}\right\}. \quad (4)$$

Lemma 2 essentially states a sufficient condition on $p$ and $T$ for $\Delta(p,T,k) \geq 0$ for any $k \in \mathbb{Z}^+$, thereby eliminating all but the two largest candidate values for $m^*$ in (3):

**Lemma 2.** *If $T \geq 2$ such that*

$$(1-p)^{\lfloor T \rfloor} + 2\lfloor T \rfloor p (1-p)^{\lfloor T \rfloor - 1} - 1 \leq 0, \quad (5)$$

*then $\bar{\mathbf{x}}(n,T,m{=}\lfloor \lfloor \tfrac{n}{T} \rfloor T \rfloor)$ or $\bar{\mathbf{x}}(n,T,m{=}n)$ is an optimal symmetric allocation.*

*Proof of Lemma 2:* Suppose that $T \geq 2$. We will show that if condition (5) is satisfied, then $\Delta(p,T,k) \geq 0$ for any $k \in \mathbb{Z}^+$. First, we note that

$$\frac{\binom{\lfloor kT \rfloor}{k-1}}{\binom{\lfloor kT \rfloor}{k}} = \frac{k}{\lfloor kT \rfloor - k + 1} = \frac{k}{\lfloor k(\lfloor T \rfloor + \tau) \rfloor - k + 1}, \quad \text{where } \tau \triangleq T - \lfloor T \rfloor \in [0,1)$$
$$= \frac{k}{k\lfloor T \rfloor + \lfloor k\tau \rfloor - k + 1} \geq \frac{k}{k\lfloor T \rfloor}, \quad \text{since } \begin{array}{l}\lfloor k\tau \rfloor \leq k\tau < k \\ \iff \lfloor k\tau \rfloor \leq k - 1 \\ \iff \lfloor k\tau \rfloor - k + 1 \leq 0\end{array}$$
$$= \frac{1}{\lfloor T \rfloor}. \quad (6)$$

Now, if condition (5) is satisfied, then

$$(1-p)^{\lfloor T \rfloor} + 2\lfloor T \rfloor p (1-p)^{\lfloor T \rfloor - 1} - 1 \leq 0$$
$$\iff \mathbb{P}[\mathcal{B}(\lfloor T \rfloor, p) = 0] + 2\mathbb{P}[\mathcal{B}(\lfloor T \rfloor, p) = 1] - 1 \leq 0$$
$$\iff \mathbb{P}[\mathcal{B}(\lfloor T \rfloor, p) \geq 2] \geq \mathbb{P}[\mathcal{B}(\lfloor T \rfloor, p) = 1]$$

$$\iff \sum_{j=2}^{\lfloor T \rfloor} \binom{\lfloor T \rfloor}{j} p^j (1-p)^{\lfloor T \rfloor - j} \geq \lfloor T \rfloor p (1-p)^{\lfloor T \rfloor - 1}$$

$$\iff \sum_{j=2}^{\lfloor T \rfloor} \frac{1}{\lfloor T \rfloor} \binom{\lfloor T \rfloor}{j} \left( \frac{p}{1-p} \right)^{j-1} \geq 1 \tag{7}$$

$$\implies \sum_{j=2}^{\lceil T \rceil} \frac{1}{\lfloor T \rfloor} \binom{\lceil T \rceil}{j} \left( \frac{p}{1-p} \right)^{j-1} \geq 1. \tag{8}$$

Observe that $\alpha_{k,T} \triangleq \lfloor (k+1)T \rfloor - \lfloor kT \rfloor \in \{\lfloor T \rfloor, \lceil T \rceil\}$, because $\alpha_{k,T} \in (T-1, T+1)$ and there are only two integers $\lfloor T \rfloor$ and $\lceil T \rceil$, which are possibly nondistinct, in this interval. It follows from (7) and (8) that

$$\sum_{j=2}^{\alpha_{k,T}} \frac{1}{\lfloor T \rfloor} \binom{\alpha_{k,T}}{j} \left( \frac{p}{1-p} \right)^{j-1} \geq 1. \tag{9}$$

Therefore, we have

$$\sum_{i=1}^{\min(\alpha_{k,T}-1, k)} \sum_{j=i+1}^{\alpha_{k,T}} \frac{\binom{\lfloor kT \rfloor}{k-i}}{\binom{\lfloor kT \rfloor}{k}} \binom{\alpha_{k,T}}{j} \left( \frac{p}{1-p} \right)^{-i+j}$$

$$\geq \sum_{i=1}^{1} \sum_{j=i+1}^{\alpha_{k,T}} \frac{\binom{\lfloor kT \rfloor}{k-i}}{\binom{\lfloor kT \rfloor}{k}} \binom{\alpha_{k,T}}{j} \left( \frac{p}{1-p} \right)^{-i+j}, \quad \begin{array}{l} \text{since} \\ \min(\alpha_{k,T}-1, k) \\ \geq \min(2-1, 1) \\ = 1 \end{array}$$

$$= \sum_{j=2}^{\alpha_{k,T}} \frac{\binom{\lfloor kT \rfloor}{k-1}}{\binom{\lfloor kT \rfloor}{k}} \binom{\alpha_{k,T}}{j} \left( \frac{p}{1-p} \right)^{j-1}$$

$$\geq \sum_{j=2}^{\alpha_{k,T}} \frac{1}{\lfloor T \rfloor} \binom{\alpha_{k,T}}{j} \left( \frac{p}{1-p} \right)^{j-1}, \text{ from (6)}$$

$$\geq 1, \text{ from (9)}$$

$$\implies \sum_{i=1}^{\min(\alpha_{k,T}-1, k)} \sum_{j=i+1}^{\alpha_{k,T}} \binom{\lfloor kT \rfloor}{k-i} \binom{\alpha_{k,T}}{j} \left( \frac{p}{1-p} \right)^{-i+j} \geq \binom{\lfloor kT \rfloor}{k}$$

$$\iff \Delta(p, T, k) \geq 0, \text{ from (4)}.$$

Thus, we conclude that $P_S(p, T, m=\lfloor T \rfloor) \leq P_S(p, T, m=\lfloor 2T \rfloor) \leq \cdots \leq P_S(p, T, m=\lfloor \frac{n}{T} \rfloor T)$, and so we can find an optimal $m^*$ from among $\{\lfloor \frac{n}{T} \rfloor T, n\}$. ∎

Theorem 2 restates Lemma 2 in a slightly weaker but more convenient form:

*Proof of Theorem 2:* Since $\left\lceil \frac{4}{3p} \right\rceil \geq \left\lceil \frac{4}{3} \right\rceil = 2$ and

$$T \geq \left\lceil \frac{4}{3p} \right\rceil \iff \lfloor T \rfloor \geq \left\lceil \frac{4}{3p} \right\rceil \geq \frac{4}{3p} \iff p \geq \frac{4}{3\lfloor T \rfloor},$$

it follows that if $T \geq \left\lceil \frac{4}{3p} \right\rceil$, then $T \geq 2$ and $p \geq \frac{4}{3\lfloor T \rfloor}$. We will show that condition (5) of Lemma 2 is satisfied for any $T \geq 2$ and $p \geq \frac{4}{3\lfloor T \rfloor}$. To do this, we define the function

$$f(p, T) \triangleq (1-p)^{\lfloor T \rfloor} + 2\lfloor T \rfloor p (1-p)^{\lfloor T \rfloor - 1} - 1,$$

and show that $f(p, T) \leq f\left( p = \frac{4}{3\lfloor T \rfloor}, T \right) \leq 0$ for any $T \geq 2$ and $p \geq \frac{4}{3\lfloor T \rfloor}$.

The partial derivative of $f(p, T)$ wrt $p$ is given by

$$\frac{\partial}{\partial p} f(p, T) = \lfloor T \rfloor (1-p)^{\lfloor T \rfloor - 2} (1 + p - 2\lfloor T \rfloor p).$$

Observe that $f(p, T)$ is decreasing wrt $p$ for any $T \geq 2$ and $p \geq \frac{4}{3\lfloor T \rfloor}$, since

$$p \geq \frac{4}{3\lfloor T \rfloor} = \frac{1}{\frac{3}{4}\lfloor T \rfloor} > \frac{1}{2\lfloor T \rfloor - 1}$$

$$\implies 2\lfloor T \rfloor p - p > 1 \iff 1 + p - 2\lfloor T \rfloor p < 0 \iff \frac{\partial}{\partial p} f(p, T) < 0.$$

Now, consider the function

$$g(T) \triangleq f\left( p = \frac{4}{3\lfloor T \rfloor}, T \right) = \left( 1 - \frac{4}{3\lfloor T \rfloor} \right)^{\lfloor T \rfloor - 1} \left( \frac{11}{3} - \frac{4}{3\lfloor T \rfloor} \right) - 1.$$

We will proceed to show that $g(T) \leq 0$ for any $T \geq 2$. For $T \in [2, 3)$, we have $\lfloor T \rfloor = 2$ and $g(T) = 0$. To show that $g(T) \leq 0$ for $T \geq 3$, consider the function

$$h(T) \triangleq (T-1) \ln \left( 1 - \frac{4}{3T} \right) + \ln \left( \frac{11}{3} - \frac{4}{3T} \right),$$

which has the derivatives

$$h'(T) = \frac{1}{3T-4} + \frac{11}{11T-4} + \ln \left( 1 - \frac{4}{3T} \right),$$

$$h''(T) = \frac{16 \left( 11T^2 - 24T - 16 \right)}{T \left( 33T^2 - 56T + 16 \right)^2}.$$

Since $h'(T=3) = \frac{84}{145} - \ln \frac{9}{5} < 0$, $\lim_{T \to \infty} h'(T) = 0$, and $h''(T) > 0$ for $T \geq 3$, it follows that $h'(T) \leq 0$ for $T \geq 3$. Now, since $h(T=3) = \ln \frac{29}{9} - 2 \ln \frac{9}{5} < 0$, and $h'(T) \leq 0$ for $T \geq 3$, it follows that $h(T) < 0$ for $T \geq 3$. Thus, for $T \geq 3$, we have

$$(\lfloor T \rfloor - 1) \ln \left( 1 - \frac{4}{3\lfloor T \rfloor} \right) + \ln \left( \frac{11}{3} - \frac{4}{3\lfloor T \rfloor} \right) = h(\lfloor T \rfloor) < 0$$

$$\iff \ln \left\{ \left( 1 - \frac{4}{3\lfloor T \rfloor} \right)^{\lfloor T \rfloor - 1} \left( \frac{11}{3} - \frac{4}{3\lfloor T \rfloor} \right) \right\} < 0$$

$$\iff \left( 1 - \frac{4}{3\lfloor T \rfloor} \right)^{\lfloor T \rfloor - 1} \left( \frac{11}{3} - \frac{4}{3\lfloor T \rfloor} \right) < 1 \iff g(T) < 0.$$

Combining these results, we obtain

$$f(p, T) \leq f\left( p = \frac{4}{3\lfloor T \rfloor}, T \right) \triangleq g(T) \leq 0$$

for any $T \geq 2$ and $p \geq \frac{4}{3\lfloor T \rfloor}$. ∎

Lemma 3 mirrors Lemma 2 by stating a sufficient condition on $p$ and $T$ for $\Delta(p, T, k) \leq 0$ for any $k \in \mathbb{Z}^+$, thereby eliminating all but the smallest candidate value for $m^*$ in (3):

**Lemma 3.** *If $T > 1$ such that either*

$$T = \frac{1}{p} \in \mathbb{Z}^+, \tag{10}$$

*or*

$$T < \frac{1}{p} \quad \text{and} \quad p(1-p)^{\lceil T \rceil - 1} \leq \frac{1}{T} \left( 1 - \frac{1}{T} \right)^{\lceil T \rceil - 1}, \tag{11}$$

*then $\bar{\mathbf{x}}(n, T, m=\lfloor T \rfloor)$ is an optimal symmetric allocation.*

*Proof of Lemma 3:* Suppose that $T > 1$. We will show that if condition (10) or condition (11) is satisfied, then $\Delta(p, T, k) \leq 0$ for any $k \in \mathbb{Z}^+$. First, we note that for any $i \in \{1, \ldots, k\}$,

$$\frac{\binom{\lfloor kT \rfloor}{k-i}}{\binom{\lfloor kT \rfloor}{k}} = \frac{\overbrace{(k)(k-1)\cdots(k-i+1)}^{i \text{ terms}}}{\underbrace{(\lfloor kT \rfloor - k+i)\cdots(\lfloor kT \rfloor - k+2)(\lfloor kT \rfloor - k+1)}_{i \text{ terms}}}$$

$$\leq \left(\frac{k}{\lfloor kT \rfloor - k+1}\right)^i \leq \left(\frac{k}{kT-1-k+1}\right)^i, \text{ since } kT-1 < \lfloor kT \rfloor$$

$$= \left(\frac{1}{T-1}\right)^i. \quad (12)$$

Now, if condition (10) is satisfied, then

$$\sum_{i=1}^{\lceil T \rceil - 1} \sum_{j=i+1}^{\lceil T \rceil} \left(\frac{1}{T-1}\right)^i \binom{\lceil T \rceil}{j} \left(\frac{p}{1-p}\right)^{-i+j}$$

$$= \sum_{i=1}^{T-1} \sum_{j=i+1}^{T} \left(\frac{1}{T-1}\right)^i \binom{T}{j} \left(\frac{\frac{1}{T}}{1-\frac{1}{T}}\right)^{-i+j}$$

$$= \sum_{i=1}^{T-1} \sum_{j=i+1}^{T} \binom{T}{j} \left(\frac{1}{T-1}\right)^j$$

$$= \sum_{\ell=2}^{T} (\ell-1) \binom{T}{\ell} \left(\frac{1}{T-1}\right)^\ell = 1.$$

On the other hand, if condition (11) is satisfied, then

$$\sum_{i=1}^{\lceil T \rceil - 1} \sum_{j=i+1}^{\lceil T \rceil} \left(\frac{1}{T-1}\right)^i \binom{\lceil T \rceil}{j} \left(\frac{p}{1-p}\right)^{-i+j}$$

$$= \sum_{i=1}^{\lceil T \rceil - 1} \sum_{j=i+1}^{\lceil T \rceil} \binom{\lceil T \rceil}{j} \left(\frac{1-p}{p(T-1)}\right)^i \left(\frac{p}{1-p}\right)^j$$

$$= \sum_{\ell=2}^{\lceil T \rceil} \left(\sum_{r=1}^{\ell-1} \left(\frac{1-p}{p(T-1)}\right)^r\right) \binom{\lceil T \rceil}{\ell} \left(\frac{p}{1-p}\right)^\ell$$

$$= 1 - \frac{T\left(\frac{1}{T}\left(1-\frac{1}{T}\right)^{\lceil T \rceil - 1} - p(1-p)^{\lceil T \rceil - 1}\right)}{(1-pT)\left(1-\frac{1}{T}\right)^{\lceil T \rceil - 1}(1-p)^{\lceil T \rceil - 1}} \leq 1.$$

Thus, if either condition is satisfied, we have

$$\sum_{i=1}^{\lceil T \rceil - 1} \sum_{j=i+1}^{\lceil T \rceil} \left(\frac{1}{T-1}\right)^i \binom{\lceil T \rceil}{j} \left(\frac{p}{1-p}\right)^{-i+j} \leq 1 \quad (13)$$

$$\implies \sum_{i=1}^{\lfloor T \rfloor - 1} \sum_{j=i+1}^{\lfloor T \rfloor} \left(\frac{1}{T-1}\right)^i \binom{\lfloor T \rfloor}{j} \left(\frac{p}{1-p}\right)^{-i+j} \leq 1. \quad (14)$$

As in the proof of Lemma 2, we note that $\alpha_{k,T} \triangleq \lfloor (k+1)T \rfloor - \lfloor kT \rfloor \in \{\lfloor T \rfloor, \lceil T \rceil\}$. It follows from (13) and (14) that

$$\sum_{i=1}^{\alpha_{k,T}-1} \sum_{j=i+1}^{\alpha_{k,T}} \left(\frac{1}{T-1}\right)^i \binom{\alpha_{k,T}}{j} \left(\frac{p}{1-p}\right)^{-i+j} \leq 1. \quad (15)$$

Therefore, we have

$$\sum_{i=1}^{\min(\alpha_{k,T}-1,k)} \sum_{j=i+1}^{\alpha_{k,T}} \frac{\binom{\lfloor kT \rfloor}{k-i}}{\binom{\lfloor kT \rfloor}{k}} \binom{\alpha_{k,T}}{j} \left(\frac{p}{1-p}\right)^{-i+j}$$

$$\leq \sum_{i=1}^{\min(\alpha_{k,T}-1,k)} \sum_{j=i+1}^{\alpha_{k,T}} \left(\frac{1}{T-1}\right)^i \binom{\alpha_{k,T}}{j} \left(\frac{p}{1-p}\right)^{-i+j}, \text{ from (12)}$$

$$\leq \sum_{i=1}^{\alpha_{k,T}-1} \sum_{j=i+1}^{\alpha_{k,T}} \left(\frac{1}{T-1}\right)^i \binom{\alpha_{k,T}}{j} \left(\frac{p}{1-p}\right)^{-i+j}, \text{ since } \min(\alpha_{k,T}-1,k) \leq \alpha_{k,T}-1$$

$$\leq 1, \text{ from (15)}$$

$$\implies \sum_{i=1}^{\min(\alpha_{k,T}-1,k)} \sum_{j=i+1}^{\alpha_{k,T}} \binom{\lfloor kT \rfloor}{k-i} \binom{\alpha_{k,T}}{j} \left(\frac{p}{1-p}\right)^{-i+j} \leq \binom{\lfloor kT \rfloor}{k}$$

$$\iff \Delta(p,T,k) \leq 0, \text{ from (4)}.$$

It follows that $P_{\text{S}}(p,T,m=\lfloor T \rfloor) \geq P_{\text{S}}(p,T,m=\lfloor 2T \rfloor) \geq P_{\text{S}}(p,T,m=\lfloor 3T \rfloor) \geq \cdots$. Since

$$P_{\text{S}}(p,T,m=n) \begin{cases} = P_{\text{S}}(p,T,m=\lfloor \lfloor \frac{n}{T} \rfloor T \rfloor) & \text{if } \frac{n}{T} \in \mathbb{Z}^+, \\ \leq P_{\text{S}}(p,T,m=\lfloor (\lfloor \frac{n}{T} \rfloor + 1)T \rfloor) & \text{otherwise}, \end{cases}$$

we conclude that $m = \lfloor T \rfloor$ gives an optimal symmetric allocation. ∎

Lemma 4 restates Lemma 3 in a slightly weaker but more convenient form:

**Lemma 4.** *If $T > 1$ and $p \leq \frac{2}{\lceil T \rceil} - \frac{1}{T}$, then $\bar{\mathbf{x}}(n, T, m=\lfloor T \rfloor)$ is an optimal symmetric allocation.*

*Proof of Lemma 4:* We will show that either condition (10) or condition (11) of Lemma 3 is satisfied for any $T > 1$ and $p \leq \frac{2}{\lceil T \rceil} - \frac{1}{T}$. We do this in two steps: First, we define the function

$$f(p,T) \triangleq \frac{p(1-p)^{\lceil T \rceil - 1}}{\frac{1}{T}\left(1-\frac{1}{T}\right)^{\lceil T \rceil - 1}} - 1,$$

and show that $f(p,T) \leq f\left(p = \frac{2}{\lceil T \rceil} - \frac{1}{T}, T\right) \leq 0$ for any $T > 1$ and $p \leq \frac{2}{\lceil T \rceil} - \frac{1}{T}$. Second, we apply the appropriate condition from Lemma 3 for each pair of $T$ and $p$.

The partial derivative of $f(p,T)$ wrt $p$ is given by

$$\frac{\partial}{\partial p} f(p,T) = \frac{(1-p\lceil T \rceil)(1-p)^{\lceil T \rceil - 2}}{\frac{1}{T}\left(1-\frac{1}{T}\right)^{\lceil T \rceil - 1}}.$$

Observe that $f(p,T)$ is nondecreasing wrt $p$ for any $T > 1$ and $p \leq \frac{2}{\lceil T \rceil} - \frac{1}{T}$, since

$$p \leq \frac{2}{\lceil T \rceil} - \frac{1}{T} \leq \frac{2}{\lceil T \rceil} - \frac{1}{\lceil T \rceil} = \frac{1}{\lceil T \rceil}$$

$$\implies p\lceil T \rceil \leq 1 \iff 1 - p\lceil T \rceil \geq 0 \iff \frac{\partial}{\partial p} f(p,T) \geq 0.$$

Now, consider the function

$$g(T) \triangleq f\left(p = \frac{2}{\lceil T \rceil} - \frac{1}{T}, T\right) = \frac{\left(\frac{2}{\lceil T \rceil} - \frac{1}{T}\right)\left(1 - \frac{2}{\lceil T \rceil} + \frac{1}{T}\right)^{\lceil T \rceil - 1}}{\frac{1}{T}\left(1 - \frac{1}{T}\right)^{\lceil T \rceil - 1}} - 1.$$

We will proceed to show that $g(T) \leq 0$ for any $T > 1$ by reparameterizing $g(T)$ as $h(c, \tau)$, where $c \triangleq \lceil T \rceil$ and $\tau \triangleq \lceil T \rceil - T$:

$$h(c, \tau) \triangleq g(T = c - \tau) = \frac{\left(\frac{2}{c} - \frac{1}{c-\tau}\right)\left(1 - \frac{2}{c} + \frac{1}{c-\tau}\right)^{c-1}}{\frac{1}{c-\tau}\left(1 - \frac{1}{c-\tau}\right)^{c-1}} - 1.$$

The partial derivative of $h(c, \tau)$ wrt $\tau$ is given by

$$\frac{\partial}{\partial \tau} h(c, \tau) = -\frac{2\tau^2(c-2)\left(1 - \frac{2}{c} + \frac{1}{c-\tau}\right)^c}{(c(c-1-\tau) + 2\tau)^2 \left(1 - \frac{1}{c-\tau}\right)^c}.$$

Since $\frac{\partial}{\partial \tau} h(c, \tau) \leq 0$ for any $c \in \mathbb{Z}^+$, $c \geq 2$, and $\tau \in [0, 1)$, it follows that for any $T > 1$, we have

$$g(T) = h\left(c = \lceil T \rceil, \tau = \lceil T \rceil - T\right)$$
$$\leq h\left(c = \lceil T \rceil, \tau = 0\right)$$
$$= \frac{\left(\frac{2}{\lceil T \rceil} - \frac{1}{\lceil T \rceil}\right)\left(1 - \frac{2}{\lceil T \rceil} + \frac{1}{\lceil T \rceil}\right)^{\lceil T \rceil - 1}}{\frac{1}{\lceil T \rceil}\left(1 - \frac{1}{\lceil T \rceil}\right)^{\lceil T \rceil - 1}} - 1 = 0.$$

Combining these results, we obtain

$$f(p, T) \leq f\left(p = \frac{2}{\lceil T \rceil} - \frac{1}{T}, T\right) \triangleq g(T) \leq 0$$

for any $T > 1$ and $p \leq \frac{2}{\lceil T \rceil} - \frac{1}{T}$, which implies

$$p(1-p)^{\lceil T \rceil - 1} \leq \frac{1}{T}\left(1 - \frac{1}{T}\right)^{\lceil T \rceil - 1}.$$

Finally, we apply the appropriate condition from Lemma 3 for each pair of $T$ and $p$. For $T \in \mathbb{Z}^+, T > 1$, we have $\frac{2}{\lceil T \rceil} - \frac{1}{T} = \frac{1}{T}$: we use condition (10) for $p = \frac{1}{T}$, and condition (11) for $p < \frac{1}{T}$. For $T \notin \mathbb{Z}^+, T > 1$, we have $\frac{2}{\lceil T \rceil} - \frac{1}{T} < \frac{1}{T}$: we use condition (11) for $p < \frac{1}{T}$. ∎

Theorem 3 expands the region covered by Lemma 4 by showing that $\bar{\mathbf{x}}(n, T, m = \lfloor T \rfloor)$ remains optimal between the "peaks" in Fig. 3:

*Proof of Theorem 3:* Since

$$T \leq \left\lfloor \frac{1}{p} \right\rfloor \iff \lceil T \rceil \leq \left\lfloor \frac{1}{p} \right\rfloor \leq \frac{1}{p} \iff p \leq \frac{1}{\lceil T \rceil},$$

it suffices to show that $\bar{\mathbf{x}}(n, T, m = \lfloor T \rfloor)$ is an optimal symmetric allocation for any $T > 1$ and $p \leq \frac{1}{\lceil T \rceil}$. (The theorem is trivially true for $T = 1$.) We do this by considering subintervals of $T$ over which $\lceil T \rceil$ is constant.

Let $T$ be confined to the unit interval $(c, c+1]$, where $c \in \mathbb{Z}^+$. According to Lemma 4, $\bar{\mathbf{x}}(n, T, m = \lfloor T \rfloor)$ is optimal for any $p \in \left(0, \frac{2}{c+1} - \frac{1}{T}\right]$ and $T \in (c, c+1]$, or equivalently, for any

$$p \in \left(0, \frac{1}{c+1}\right] \quad \text{and} \quad T \in \left[\frac{1}{\frac{2}{c+1} - p}, c+1\right] \cap (c, c+1].$$

This is just the area below a "peak" in Fig. 3, expressed in terms of different independent variables. For each $p \in \left(0, \frac{1}{c+1}\right)$, we can always find a $T_0$ such that $T_0 \in \left[\frac{1}{\frac{2}{c+1} - p}, c+1\right) \cap (c, c+1)$; for example, we can pick $T_0 = c + 1 - \delta$, where $\delta \triangleq \frac{1}{2}\left(c + 1 - \max\left(c, \frac{1}{\frac{2}{c+1} - p}\right)\right) \in (0, 1)$. Now, we make the crucial observation that if $\bar{\mathbf{x}}(n, T, m = \lfloor T \rfloor)$ is an optimal symmetric allocation for $T = T_0$, then $\bar{\mathbf{x}}(n, T, m = \lfloor T \rfloor)$ is also an optimal symmetric allocation for any $T \in [\lfloor T_0 \rfloor, T_0]$. This claim can be proven by contradiction: the recovery probability for $\bar{\mathbf{x}}(n, T, m = \lfloor T \rfloor)$ is $P_S(p, T, m = \lfloor T \rfloor) = \mathbb{P}[\mathcal{B}(\lfloor T \rfloor, p) \geq 1]$ which remains constant for all $T \in [\lfloor T_0 \rfloor, T_0]$, and a symmetric allocation that performs strictly better than $\bar{\mathbf{x}}(n, T, m = \lfloor T \rfloor)$ for some $T \in [\lfloor T_0 \rfloor, T_0]$ would therefore also outperform $\bar{\mathbf{x}}(n, T, m = \lfloor T \rfloor)$ for $T = T_0$. Since $\bar{\mathbf{x}}(n, T, m = \lfloor T \rfloor)$ is indeed optimal for our choice of $T_0$, it follows then that $\bar{\mathbf{x}}(n, T, m = \lfloor T \rfloor)$ is also optimal for any $p \in \left(0, \frac{1}{c+1}\right)$ and $T \in (c, c+1]$. By applying this result for each $c \in \mathbb{Z}^+$, we reach the conclusion that $\bar{\mathbf{x}}(n, T, m = \lfloor T \rfloor)$ is an optimal symmetric allocation for any $T > 1$ and $p < \frac{1}{\lceil T \rceil}$.

Finally, to extend the optimality of $\bar{\mathbf{x}}(n, T, m = \lfloor T \rfloor)$ to $p = \frac{1}{\lceil T \rceil}$, we note that the recovery probability $P_S(p, T, m) \triangleq \mathbb{P}\left[\mathcal{B}(m, p) \geq \left\lceil \frac{m}{T} \right\rceil\right]$ is a polynomial in $p$ and is therefore continuous at $p = \frac{1}{\lceil T \rceil}$. Since $\bar{\mathbf{x}}(n, T, m = \lfloor T \rfloor)$ is optimal as $p \to \frac{1}{\lceil T \rceil}^{-}$, it remains optimal at $p = \frac{1}{\lceil T \rceil}$. ∎

## REFERENCES

[1] R. Karp, R. Kleinberg, C. Papadimitriou, and E. Friedman, *Personal communication*, 2006.
[2] S. Jain, M. Demmer, R. Patra, and K. Fall, "Using redundancy to cope with failures in a delay tolerant network," in *Proc. ACM SIGCOMM*, Aug. 2005.
[3] D. Leong, A. G. Dimakis, and T. Ho, "Distributed storage allocation problems," in *Proc. Workshop Netw. Coding, Theory, and Appl. (NetCod)*, Jun. 2009.
[4] ——, "Distributed storage allocation for high reliability," in *Proc. IEEE Int. Conf. Commun. (ICC)*, May 2010.
[5] M. Sardari, R. Restrepo, F. Fekri, and E. Soljanin, "Memory allocation in distributed storage networks," in *Proc. IEEE Int. Symp. Inf. Theory (ISIT)*, Jun. 2010.
[6] M. Naor and R. M. Roth, "Optimal file sharing in distributed networks," *SIAM J. Comput.*, vol. 24, no. 1, pp. 158–183, Feb. 1995.
[7] A. Jiang and J. Bruck, "Network file storage with graceful performance degradation," *ACM Trans. Storage*, vol. 1, no. 2, pp. 171–189, May 2005.
[8] M. Mitzenmacher and E. Upfal, *Probability and Computing: Randomized Algorithms and Probabilistic Analysis*. Cambridge University Press, 2005.